\documentclass{emulateapj}
\usepackage{amssymb}
\usepackage{natbib}
\usepackage{graphicx}
\usepackage{lscape}

\def\N1Mpc{$N_{\rm 1Mpc}$\/ }
\def\L200{$L_{200}$\/}
\def\n200{$N_{200}^{\rm gal}$\/}
\def\LBCG{$L_{\rm BCG}$\/}

\def\ltsima{$\; \buildrel < \over \sim \;$}
\def\simlt{\lower.5ex\hbox{\ltsima}}
\def\gtsima{$\; \buildrel > \over \sim \;$}
\def\simgt{\lower.5ex\hbox{\gtsima}}
\def\simless{\mathbin{\lower 3pt\hbox
   {$\rlap{\raise 5pt\hbox{$\char'074$}}\mathchar"7218$}}}   
\def\simgreat{\mathbin{\lower 3pt\hbox
   {$\rlap{\raise 5pt\hbox{$\char'076$}}\mathchar"7218$}}}   

\long\def\symbolfootnote[#1]#2{\begingroup%
\def\thefootnote{\fnsymbol{footnote}}\footnote[#1]{#2}\endgroup} 

\shorttitle{Physical Properties of Four SZE-Selected Clusters}
\shortauthors{Menanteau \& Hughes}
\submitted{accepted for publication to ApJL}

\begin{document}

\title{Physical Properties of Four SZE-Selected Galaxy Clusters in the 
Southern Cosmology Survey}

\author{Felipe Menanteau and John P. Hughes}
\affil{Rutgers University, Department of Physics \& Astronomy, 136 
Frelinghuysen Road, Piscataway, NJ 08854, USA }

\begin{abstract}

We present the optical and X-ray properties of four clusters recently
discovered by the South Pole Telescope (SPT) using the
Sunyaev-Zel'dovich effect (SZE).  The four clusters are located in one
of the common survey areas of the southern sky that is also being
targeted by the Atacama Cosmology Telescope (ACT) and  imaged
by the CTIO Blanco 4-m telescope. Based on publicly available $griz$
optical images and {\it XMM-Newton} and {\it ROSAT} X-ray observations
we analyse the physical properties of these clusters and obtain
photometric redshifts, luminosities, richness and mass estimates.
Each cluster contains a central elliptical whose luminosity is
consistent with SDSS cluster studies.  Our mass estimates are well
above the nominal detection limit of SPT and ACT; the new SZE clusters
are very likely massive systems with $M\simgt 5\times 10^{14}\,M_\sun$.

\end{abstract}


\section{Introduction}

Galaxy clusters serve as important cosmological probes as their
formation and evolution rate depend on cosmological parameters and the
kinematics of the dark matter. They are the observational counterparts
of dark matter halos and their abundances and masses as a function of
redshift are quite sensitive to the growth of structure in the
Universe offering a potentially powerful probe of dark energy
\citep{Carlstrom-02}. Galaxy clusters also harbor a significant
fraction of the visible baryons in the Universe. Hydrogen gas, heated
as it falls into the dark-matter potential well of a cluster, forms an
intracluster medium too hot to be bound by individual galaxies. The
hot intracluster gas leaves an imprint on the Cosmic Microwave
Background Radiation (CMBR) though the Sunyaev Zel'dovich effect (SZE)
\citep{SZ72} in which CMB photons are inverse-Compton scattered by the
hot intracluster gas.  It is this SZE signal that two new ground-based
mm-band telescopes, the Atacama Cosmology Telescope (ACT)
\citep{Kosowsky06,Fowler07} and the South Pole Telescope (SPT)
\citep{SPTref}, have been designed to detect through its
frequency-dependence: these experiments will measure intensity shifts
shifts of the CMB radiation corresponding to a decrement below and an
increment above the ``null'' frequency around 220~GHz. Both telescopes
are now actively acquiring data over large areas of the southern sky
that will ultimately cover several hundreds to thousands of square
degrees.

Recent results from the SPT collaboration \citep{SPT1} have yielded
the first blind detection of SZE clusters in one of the common
southern survey regions centered near right ascension 05$^{\rm
  hr}$30$^{\rm m}$ and declination $-$53$^\circ$. This area has also
been scanned by ACT and has been optically imaged in the optical
($griz$) with the CTIO Blanco 4-m telescope as part of the Blanco
Cosmology Survey (BCS) \citep{SCSI,SPT1}. Prompted by these results,
here we present a detailed study of the physical properties of these
first four SZE-detected clusters based on publicly available optical
imaging and archival X-ray data.  Throughout this paper we assume a
flat cosmology with $H_0=100 h$~km~s$^{-1}$~Mpc$^{-1}$, $h=0.7$ and
matter density $\Omega_M=0.3$.

\section{SZE Optical Counterparts}

\begin{figure*}
\centerline{
\includegraphics[width=7.1in]{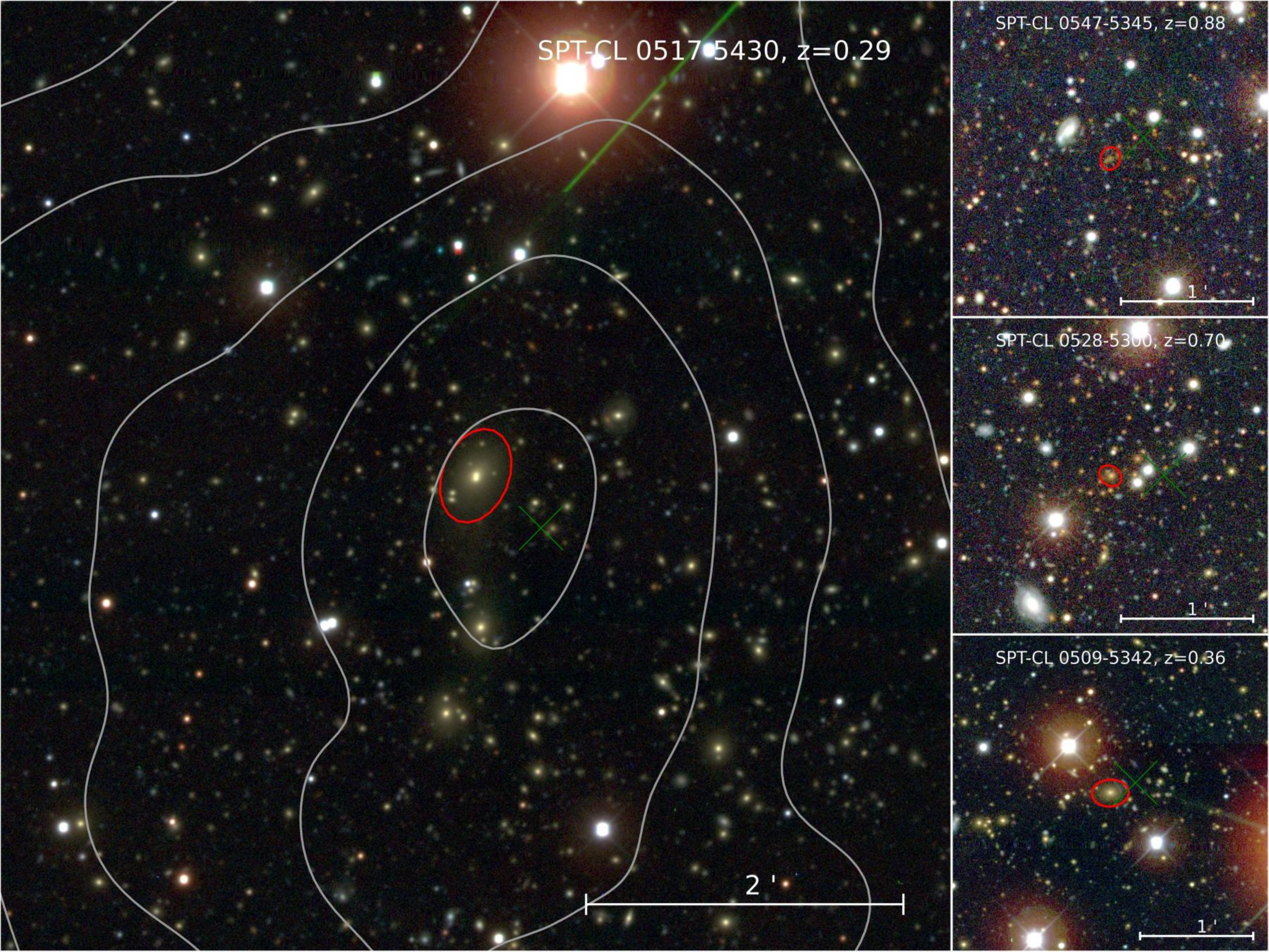}
}
\caption{Composite $gri$ color image of the four SPT SZE-selected
  clusters from the data. Green crosses show the position of the SZE
  decrement from \cite{SPT1} for each system. The red ellipses denote
  the position of the BCG and show as well the SExtractor Kron
  radius. For cluster SPT-CL~0517$-$5430 (Abell 0520, main panel) we show
  the {\it XMM-Newton} X-ray isointensity contour levels (starting
  near the center at a value of $1\times 10^{-4}$ cts s$^{-1}$
  pixel$^{-1}$ and decreasing going outward by factors of 2.5).  }
\label{fig:cplate}
\end{figure*}

The positions of the four SZE-selected clusters reported by the SPT
team are contained within the region surveyed by the BCS in 2005 and
2006. These observations have been publically available for about
a year or more now.  We have processed and analyzed them using
an independent software pipeline developed by us at Rutgers
University.  We refer the reader to \cite{SCSI} where we provide a
full description of the observing strategy, pipeline and data products
and some results on new massive galaxy clusters discovered in the BCS
survey field centered near 23hr.  Here we use the identical software
pipeline to generate data products from which we characterize the
optical properties of the four SZE clusters.  First, we identify the
optical counterparts, which appear as clearly visible overdensities of
early-type galaxies near the sky locations of the SZE decrements (see
Table~1 from \citealt{SPT1}). Roughly centered within each overdensity
is a bright elliptical, which we take to be the Brightest Cluster
Galaxy (BCG).  In Figure~\ref{fig:cplate} we show the $gri$ color
images of the four systems, centered on the location of the BCG.  Each
system shows a dominant population of early-type galaxies with very
similar colors that can easily be identified by visual inspection. In
all cases the offset between the BCG and the SZE decrement, shown as
red ellipses and green crosses respectively, is less than
$\sim30$~arcsec. 

\subsection{Photometric Redshift Determination}

We determine photometric redshifts from the four-band optical images
using BPZ \citep{BPZ} following the same procedure as in
\cite{SCSI}. To avoid contamination by non-cluster members we use the
BCG to estimate the photometric redshift of each system. The BCG is,
by definition, indisputably part of the cluster, resides in a
quasi-central location and always provides the strongest signal and
the best colors since such galaxies are generally several times
brighter than $L^*$ at any given redshift. Moreover, the luminosities
and particularly the colors of BCGs are well constrained as they are
very bright elliptical galaxies dominated by old metal-rich stellar
populations for which multi-color spectral energy distribution (SED)
fitting techniques are well suited yielding precise and reliable
results. In the bottom panel of Figure~\ref{fig:mag-z} we show the
output BPZ photometric redshift probability density function (pdf) for
each BCG. The differences between the BPZ Bayesian and Maximum
Likelihood (ML) predicted redshifts are small ($\delta z\simeq0.01$)
in all cases and therefore we will employ ML photometric redshifts
hereafter as the addition of a prior is not justified when, as is the
case here, the BCG SEDs are fitted unambiguously. We note that
occasionally BCGs host radio-loud active galactic nuclei
\citep{best07}. Although we have no indication of this, if any of our
BCGs do harbor AGN then we would expect some blueing of the colors
causing our photo-z's to be slightly over-estimated.

Only one of the clusters, SPT-CL~0517$-$5430, has been previously
reported; it is the optically-rich cluster Abell S0520 \citep{Abell89}
which is also the X-ray cluster RXC~J0516.6$-$5430 \citep{REFLEX}.
Its published spectroscopic redshift of $z=0.294$ \citep{Guzzo99} is
in excellent agreement with our photometric redshift for that BCG,
$z=0.27 \pm 0.02$, but somewhat different from the value of $z=0.35$
quoted by \cite{SPT1} for the cluster which came from the peak of
their photo-z distribution of red-sequence galaxies using the same BCS
imaging.
In Table~\ref{tab} we present the clusters' optical positions based on
the BCG locations and photometric redshifts with 1-$\sigma$
uncertainties.

\subsection{BCG luminosities}

\begin{figure}
\centerline{
\includegraphics[width=4.0in]{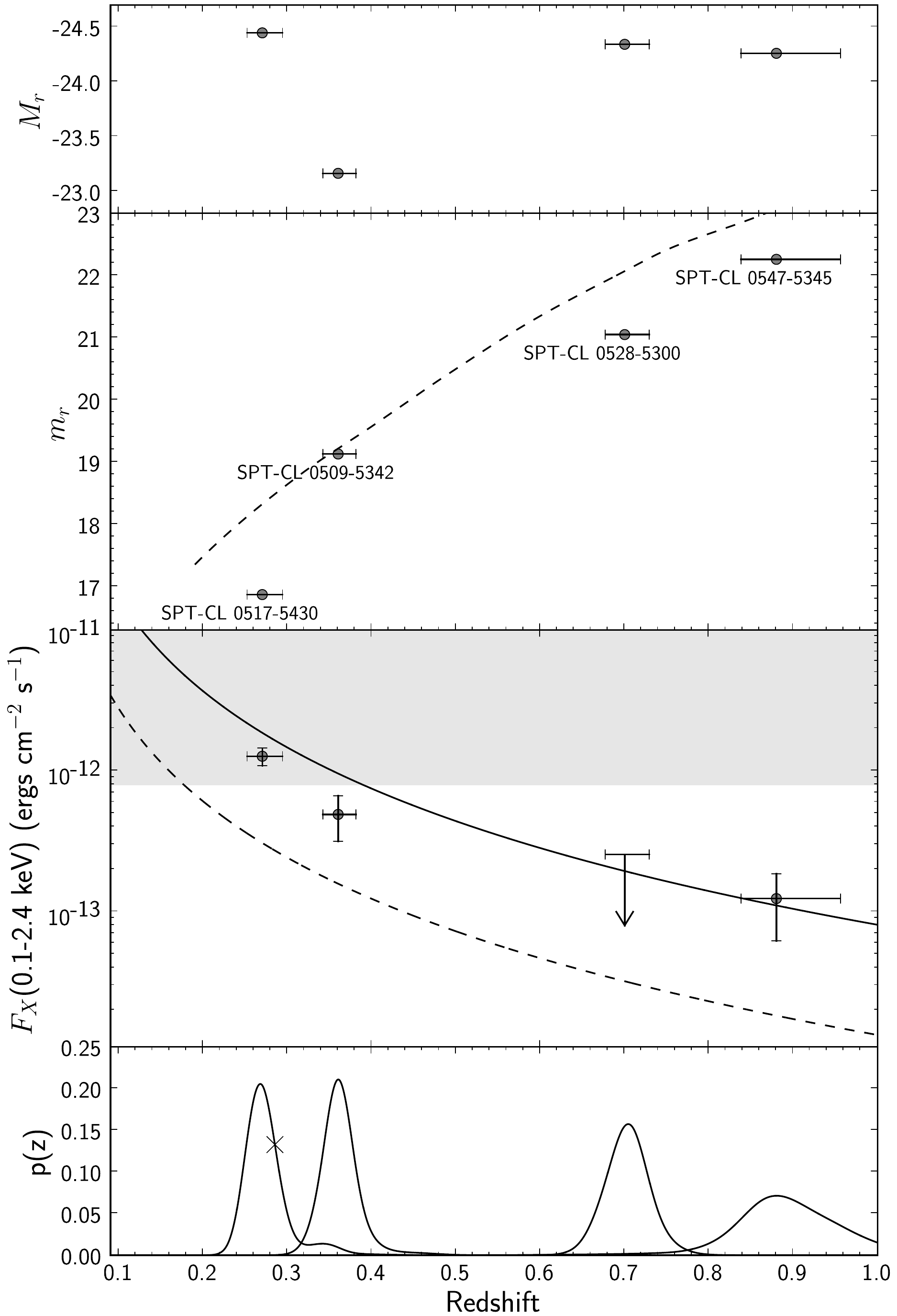}}
\caption{The absolute (upper panel) and observed (2nd panel) $r$-band
  magnitude of each BCG as a function of their photometric
  redshift. The dashed line represents the modeled observed $r$-band
  magnitude of BCGs according to \cite{LS06}. In the third panel we
  show the clusters soft X-ray flux as a function of redshift. The
  solid and dashed curves represent the expected X-ray fluxes for
  clusters with mass of $M_{200}=1\times10^{15}\,M_{\sun}$ and
  $3\times10^{14}\,M_{\sun}$ respectively. The bottom panel shows the
  photo-z probability distribution function for each BCG and with a
  cross at the location in the curve for SPT-CL~0517--5430's spectroscopic
  redshift. Error bars in the figure represent $\pm1\sigma$ deviations
  drawn from the probability functions.}
\label{fig:mag-z}
\end{figure}

The range in brightness of BCGs has been observationally established
from galaxy clusters in the SDSS up to redshifts of $0.36$ \citep[see,
for example,][]{LS06,MaxBCG}.  This can be used to determine whether
the BCGs we identify in the SZE clusters have luminosities consistent
with SDSS BCGs. In Figure~\ref{fig:mag-z} (2nd panel from top) we
show the total $r$-band observed magnitude of each cluster's BCG as a
function of redshift compared to a parametrization of
the observed $r$-band magnitudes of SDSS BCGs (dashed line)
\cite[Figure~7 in][]{LS06}.  The curve corresponds to the simple
prescription $M^*-1.5$ where $M^*$ is taken from \cite{Blanton03} and
the model includes passive evolution with redshift.
We have extrapolated the SDSS BCG luminosity model beyond $z=0.36$ as
there is overwhelming evidence that ellipticals in clusters have
passively evolved since $z\simeq1$ \citep{Blakeslee03, Lidman08,
Mei08}. We see from the figure that in all cases the BCGs in the SZE
clusters are below the model curve (i.e., are intrinsically brighter)
and we take this as strong evidence in support of them being bona fide
BCGs. It is worth noting that SPT-CL~0509$-$5342 at $z=0.36$ is the least
luminous of the four, just slightly brighter than the luminosity
predicted by the model. As we will see in the next section this
cluster also has a lower mass and X-ray luminosity compared to the
other systems. In Figure~\ref{fig:mag-z} we also show the range of
absolute magnitudes of the BCGs as a function of redshift. Here we
again see that SPT-CL~0509$-$5342 is the least luminous object, $\simeq1$
magnitude fainter than the other BCGs which all have about the same
luminosity.

\section{Cluster Mass Estimations}

The new SZE experiments aim to deliver mass-selected samples of galaxy
clusters out to large redshifts which can be used to probe the growth
of structure in the Universe and to study cluster physics. Current
N-body plus hydrodynamics simulations \citep{Motl05,Nagai06,Sehgal07}
predict robust scaling relations between the SZE signal (i.e., the
$y$-distortion due to inverse Compton scattering) and cluster
mass. However, a fundamental step before using SZE clusters as
cosmological probes is to observationally establish the relationship
between $y$ and mass using mass estimates independent of the SZE
measurements.  SZE-selected clusters will almost surely be the focus of
intense efforts to obtain mass estimates using X-ray images and
spectra, weak and strong lensing and velocity dispersions. In the
following, we present the first independent mass estimates for these
new SZE-selected clusters from archival optical and X-ray data.

\subsection{Optical Cluster Masses}

Here we use optically observed parameters of galaxies to predict the
cluster mass from weak lensing richness-mass scaling relations
established from the SDSS \citep{Johnston07,Reyes08}.  We apply the
state-of-the-art mass tracer \citep{Reyes08} built from the SDSS
maxBCG cluster sample \citep{MaxBCG} of $\simeq13,000$
optically-selected clusters. 
Here we briefly describe our procedure and refer the reader
to our earlier study \citep{SCSI} for complete details.
The quantities, \n200, \L200, and \LBCG, are the observational values
needed as input to the cluster mass scaling relation.  The cluster
richness, \n200, is the number of E/S0 galaxies within $R_{200}$ with
colors and luminosities that satisfy specific conditions for
membership. $R_{200}$ is the estimated cluster size defined as the
radius where the cluster galaxy density is $200\Omega_m^{-1}$ times
the mean space density of galaxies in the present Universe. Similarly,
$L_{200}$ is the total rest-frame integrated $r$-band luminosity of
all member galaxies included in \n200 and $L_{\rm BCG}$ is the
rest-frame $r$-band luminosity of the BCG. \cite{Reyes08} provide
power-law functions for both the luminosity-mass and richness-mass
relations (see section 5.2.1 in their paper), although here we only
use the relation based on luminosity, which should be more robust than
that based on richness.
In Table~\ref{tab} we present the cluster mass estimates and
luminosities where $M(L_{200})$ is the mass observational equivalent
of $M_{200\bar{\rho}}$\footnote{$M_{200\bar{\rho}}$ is the halo mass
  enclosed within a radius of spherical volume within which the mean
  density is 200 times the critical density.}.
For the two higher redshift clusters,
SPT-CL~0528$-$5300 at $z=0.70$ and SPT-CL~0547$-$5345 at $z=0.88$ we quote
lower-limit estimates as some fraction of low luminosity galaxies
fall below our magnitude-limit ($i\simeq22.5$). If we
make the assumption that the two higher redshift clusters have 
luminosity functions that are similar to SPT-CL~0517$-$5430 at $z=0.29$,
then we estimate that we are missing 15\% and 58\% of
the luminosity for SPT-CL~0528$-$5300 and SPT-CL~0547$-$5345 respectively. The
mass estimates in Table~\ref{tab} include these corrections.

\begin{center}
\begin{deluxetable*}{rrrrrrrrrr}
\tablecaption{Cluster Optical/X-ray Properties and Mass Estimates}
\tablewidth{0pt}
\tablehead{
\multicolumn{5}{c}{} & \multicolumn{1}{c}{$[10^{12}L_{\sun}]$} &
\multicolumn{1}{c}{$[10^{10}L_{\sun}]$} &
\multicolumn{1}{c}{[$10^{15}M_{\sun}$]} &
\multicolumn{1}{c}{[$10^{44}$~erg~s$^{-1}$]}  &
\multicolumn{1}{c}{[$10^{15}M_{\sun}$]} \\
\colhead{ID} & 
\colhead{R.A.} &
\colhead{DEC.} &
\colhead{$z_{\rm photo}$} & 
\colhead{\n200} &
\colhead{$L_{200}$} &
\colhead{$L_{\rm BCG}$} &
\colhead{$M(L_{200})$} &
\colhead{$L_{\rm X}$(0.1--2.4~keV)} &
\colhead{$M(L_{\rm X})$} 
}
\startdata
SPT-CL~0517$-$5430 & 05:16:37.4 & -54:30:01.5 & $0.27_{-0.02}^{+0.02}$ & $168.9\pm 15.3$ & $ 6.53\pm 0.15$ & $31.20$ & $1.7$     &$ 3.5\pm 0.5$  &   $0.8$ \\
SPT-CL~0509$-$5342 & 05:09:21.4 & -53:42:12.3 & $0.36_{-0.02}^{+0.02}$ & $ 76.1\pm  9.2$ & $ 2.24\pm 0.05$ & $ 9.07$ & $0.4$     &$ 2.2\pm 0.8$  &   $0.6$ \\
SPT-CL~0528$-$5300 & 05:28:05.3 & -52:59:52.8 & $0.70_{-0.02}^{+0.03}$ & $ 69.3\pm  9.8$ & $ 8.62\pm 1.20$ & $23.00$ & $\geq2.1$ &$<5.5       $  &   $<1.1$ \\
SPT-CL~0547$-$5345 & 05:46:37.7 & -53:45:31.1 & $0.88_{-0.04}^{+0.08}$ & $ 12.7\pm  3.7$ & $ 1.99\pm 0.25$ & $17.90$ & $\geq0.4$ &$ 4.7\pm 2.3$  &   $1.0$ \\
\enddata
\label{tab}
\tablecomments{Physical properties of SZE selected clusters in the SCS
  regions. Redshifts represent the photometric redshift from the
  bright elliptical in the center of the cluster with $\pm1\sigma$
  limits. The cluster position is based on the BCG.}
\end{deluxetable*}
\end{center}

\subsection{X-ray Properties and Masses}

The nearby cluster SPT-CL~0517$-$5430 has accurate measurements of its
X-ray luminosity ($L_{\rm bol} = 9.2\pm1.2 \times 10^{44}$ ergs
s$^{-1}$) and temperature ($kT = 7.5\pm0.3$ keV) from {\it XMM-Newton}
observations \citep{REFLEX2}.  There is also a hydrostatic mass
estimate of $M_{500} = 6.4\pm2.1\times 10^{14}\, M_\odot$ (note
this is at an overdensity of 500 times the critical density).
The cluster's X-ray morphology is strongly elongated: it has an axial
ratio of 1.4--1.7 with its major axis aligned $\sim$14$^\circ$ from
north (see Fig.~\ref{fig:cplate}).

We utilize the {\it ROSAT} All Sky Survey (RASS) to obtain X-ray
information on the other three clusters.  As noted by \cite{SPT1},
faint RASS sources lie close to the locations of SPT-CL~0509$-$5342
and SPT-CL~0547$-$5300.  For purposes of determining X-ray fluxes and
luminosities we will assume that these RASS sources are indeed the
X-ray counterparts to the SZ clusters. Starting with the raw X-ray
photon event lists and exposure maps from the MPE ROSAT
Archive\footnote{ftp://ftp.xray.mpe.mpg.de/rosat/archive/} we
determined count rates for the band covering PI channels 52 to 201.
Extraction regions were optimized to obtain all detected source
photons (a radius of 3$^\prime$ was sufficient) and, for the
background estimation, to reduce statistical fluctuations (using an
annular region between 5$^\prime$ and 25$^\prime$).  The X-ray source
position was used for the two SZ clusters with apparent counterparts;
for SPT-CL~0528$-$5300, the quoted SZ position was used to derive
upper limits. 

The background-subtracted count rates (SPT-CL~0509$-$5342:
0.025~cts/s; SPT-CL~0528$-$5300: $<0.015$~cts/s; SPT-CL~0547$-$5345:
0.007~cts/s) were converted to fluxes assuming a hard thermal spectrum
($kT\sim5$~keV) and accounting for each cluster's Galactic absorbing
column density \citep{NH}; values fell in the range $6.1\times10^{20}$
atoms cm$^{-2}$ to $7.8\times10^{20}$ atoms cm$^{-2}$ for all sources.
A $k$-correction was also applied to convert fluxes to the standard
0.1--2.4 keV ROSAT band in the rest frame of each cluster.  These
fluxes are plotted in Fig.~\ref{fig:mag-z} (2nd panel from bottom).
The upper limit for SPT-CL~0528$-$5300 is at 2-$\sigma$ and the soft
0.1--2.4 keV band flux for SPT-CL~0517$-$5430 was converted from the
bolometric luminosity quoted by \citet{REFLEX2}. We denote the region
(at the high flux end) that corresponds approximately to the Bright
Source Catalog from the RASS.  Also shown are curves of the expected
X-ray fluxes for clusters with masses of $M_{200} = 1\times 10^{15}\,
M_\odot$ (solid) and $3\times 10^{14}\, M_\odot$ (dotted). These
curves assume the low redshift luminosity-mass (specifically
$L_X$(0.1-2.4~keV) vs.~$M_{200}$) relation from \citet{HIFLUGCS} and,
for simplicity, no redshift evolution. This scaling relation is also
used to derive an X-ray mass estimate from the soft X-ray luminosity.
Numerical values for these quantities are presented in the last two
columns of Table~\ref{tab}.

\section{Summary}

We have established that the central ellipticals associated with the
four SZE clusters have luminosities consistent with those of BCGs from
the SDSS.
We have also determined cluster masses using two different mass
estimators based on the luminosity of the optical galaxies and the
X-ray--emitting gas. These mass tracers are reasonably consistent
within the uncertainties given by the scatter in the X-ray
luminosity-mass relation ($\sim$0.2 in the log) and the estimated
factor of 2 uncertainty in the mass derived from the optical
luminosity of the galaxies \citep{SCSI}.  Based on this evidence, all
four clusters are fairly massive ($M \simgt 5\times 10^{14}\,
M_\odot$) and therefore well above the nominal detection limit of both
SPT and ACT \citep[e.g.,][]{Sehgal07}.  Although this is an important
step in identifying the optical and X-ray counterparts to the SZ
clusters, it cannot be considered definitive. In addition the RASS
count rates could well be contaminated by X-ray emission from an AGN
or other unrelated source.  Thus the conservative reader could
conclude that the X-ray luminosities and derived masses for the three
higher redshift SPT sources represent upper limits to their true
values.  Approved {\it Chandra} observations will within the next year
resolve any doubts about the X-ray emission of these clusters.  Deeper
optical imaging and spectroscopy should also be pursued in order to
better constrain the masses of these clusters.

\acknowledgments

We would like to thank the BCS team for the planning and execution of
the CTIO Blanco observations that were used in this paper.
We have made use of the ROSAT Data Archive of the Max-Planck-Institut
f{\"u}r extraterrestrische Physik (MPE) at Garching, Germany as well as
results obtained from the High Energy Astrophysics Science Archive
Research Center (HEASARC), provided by NASA's Goddard Space Flight
Center.
Financial support was provided by the National Science Foundation
under the PIRE program (award number OISE-0530095).
JPH would like to thank finalcial support from XMM grants NNX08AX72G
and NNX08AX55G.

\end{document}